\documentclass{article}
\usepackage{graphicx} 
\usepackage{amsmath}
\usepackage{caption}
\usepackage{textcomp}
\usepackage{subcaption}
\usepackage{amssymb}
\usepackage{geometry}
\usepackage{xcolor}
\usepackage{hyperref}
\usepackage{url}
\usepackage[utf8]{inputenc}
\usepackage{authblk}

\title{Scalar perturbations on the normal and self-accelerating branch of a DGP brane and $\sigma_8$}
\author[1]{Maribel Hernández-Márquez\thanks{maribel.hernandez@nucleares.unam.mx}}
\author[2]{Celia Escamilla-Rivera \thanks{.}}
\affil[1]{Instituto de Ciencias Nucleares, Universidad Nacional Aut\'{o}noma de M\'{e}xico\\
Circuito Exterior C.U., A.P. 70-543, M\'exico D.F. 04510, M\'{e}xico}
\affil[2]{Royal Astronomical Society. Burlington House Piccadilly, London W1J 0BQ, United Kingdom.}

\date{}
\begin{document}
\maketitle

\begin{abstract}
In this work, we constrain the value of $\sigma_8$ for the normal and self-accelerating branch of a DGP brane embedded in a five-dimensional Minkowski space-time. For that purpose, we first constrain the background model parameters $H_0$, $\Omega_{m0}$, $\Omega_{r0}$ and $M$ using the Pantheon+ catalogue and a mock catalogue of gravitational waves. Then, we numerically solve the equation for dark matter scalar perturbations using the dynamic scaling solution for the master equation. Finally, we found that the evolution of matter density perturbations in both branches is different from the $\Lambda$CDM model and that the value of $\sigma_8=0.774\pm0.027$ for the normal branch and $\sigma_8=0.913\pm0.032$ for the self-accelerating branch.
\end{abstract}


\section{Introduction}
The Dvali-Gabadadze-Porrati (DGP) model considers that our universe is a 4-dimensional brane embedded in a five-dimensional Minkowski space-time \cite{dvali20004d}. There is a crossover scale $r_c$ where the 4-d gravitational potential changes to a 5-d potential. Depending on how the brane is embedded, there are two different cosmological solutions known as the self-accelerating and the normal branch. In the self-accelerating branch, it is possible to obtain an accelerated expansion if $r_c\thicksim H_0^{-1}$ \cite{PhysRevD.65.044023}. In contrast, in the normal branch, it is necessary to add some dark energy, like Holographic Dark Energy \cite{wu2008dynamics}, Agegraphic Dark Energy \cite{farajollahi2013cosmological}. However, in the normal branch, the simplest way to obtain cosmic acceleration is to consider the tension of the brane, which acts like a cosmological constant. This model is known as the $\Lambda$DGP model \cite{DGPnumericsolutionscardoso}.\\ These models: the self-accelerating branch and the $\Lambda$DGP model have been proposed as solutions to explain the enigma of dark energy and have been tested by different observations \cite{fairbairn2006supernova}, \cite{maartens2006observational},\cite{lazkoz2007cosmological},\cite{lombriser2009cosmological}. Still, studies have focused on the background dynamics. However, as it is pointed out in \cite{maartens2006observational}, it is necessary to test the DGP model with observations that involve structure formation and this requires the study of the evolution of density perturbations. But in most previous works, the constraints found using density perturbations neglected the 5D perturbations or used different approximations for the evolution of density perturbations \cite{lombriser2009cosmological}, \cite{song2007large},\cite{koyama2006structure}.\\
From the observational landscape, one observation that is related to density perturbations is the value of matter density fluctuation $\sigma_8$. Currently, there is a discrepancy between the values obtained by Redshift Space Distortions (RSD)  weak, gravitational lensing \cite{heymans2021kids}   and by the cosmic microwave radiation \cite{aghanim2020planck} for the $\Lambda$CDM model known as the $\sigma_8$ tension. In general, the $\sigma_8$ value obtained from large-scale structure observations is lower than that obtained by CMB observations in the $\Lambda$CDM model.
Therefore assessing whether modified gravity models such as DGP can alleviate this tension is a matter of considerable interest.\\
It is important to mention that DGP models have been experiencing issues in one of their branches, e.g. the existence of a ghost solution in the self-accelerating branch \cite{Gorbunov:2005zk}, however the normal branch \textbf{is free from this issue}. Furthermore the flat self-accelerating branch has been found to be excluded by CMB data and supernova luminosity distances\cite{Song:2006jk}.
Despite these challenges, we include the flat self-accelerating branch in our analysis, since it still provides a valuable framework to explore modifications to gravity at large scales and their impact on cosmological perturbations and structure growth.
Updating the observational constraints on DGP cosmologies using recent datasets, such as the latest Pantheon+ catalog of supernovae and new mock gravitational wave data, is important because observational precision has improved significantly. These new datasets enable robust tests of the DGP model's predictions for both matter density perturbations (e.g., the $\sigma_8$ parameter) and the background expansion history, which might shed light on discrepancies affecting the $\Lambda$CDM model, such as the Hubble and $\sigma_8$ tensions.\\
Therefore, in this work, we constrain the value of $\sigma_8$ for the self-accelerating and the normal branch, as far as we know, this has not been done before. This work is unique in that it constrains $\sigma_8$ separately for both the normal and self-accelerating branches using current data, which had not been done previously. Providing such updated constraints clarifies whether the DGP model or its variations can still be competitive or complementary alternatives to standard cosmology \cite{PhysRevD.80.043001} or whether their distinctive signatures in observables such as gravitational wave propagation distances can be exploited to test fundamental physics beyond $\Lambda$CDM. Hence, the significance of this work lies not just in revisiting an old model but in leveraging modern data to refine, test, and possibly extend our understanding of cosmic acceleration and structure formation beyond the standard paradigm \cite{wu2025comparison}. Furthermore, we addressed this by incorporating a modified background parametrization consistent with Planck constraints and ensuring that the initial conditions selected for perturbations are aligned with standard early-Universe physics, as constrained by CMB observations. While we do not claim to fully solve the known tension between the DGP model and CMB data; our goal is to explore whether the $\sigma_8$ tension could be alleviated within the DGP framework by leveraging its modified growth of structure.
In this direction, this work differs from previous studies in that we combine the DGP models with a data-driven MCMC analysis using supernovae, GW baselines and $f\sigma_8$ catalogues to explore the parameter degeneracies and compare them systematically against the standard cosmological model baselines.\\
 In order to constrain $\sigma_8$, in Section \ref{sec:scalarperturbations}, we derive the scalar perturbation equations for both the self-accelerating and normal branches. In these cosmological scenarios, the evolution of scalar perturbations is described by a master variable $\Omega$ that satisfies a differential partial equation known as the master equation that depends on the fifth dimension \cite{Mukohyama}. Therefore, to study the evolution of density perturbations, it is necessary first to solve the master equation.\\ In the last years, there have been different approaches to solve it, the best known are the quasi-static approximation \cite{koyama2006structure}, the dynamical scaling solution \cite{nearhorizon}, \cite{largescalesong} and the numerical solution  \cite{DGPnumericsolutionscardoso}. The quasi-static approximation has been compared with the numerical solution, and the results show that the relative error in the prediction for the evolution of matter density perturbations $\triangle_m$ is $(<4\%)$ on all scales \cite{DGPnumericsolutionscardoso}. While the value of $\Phi_{+}=\frac{\Phi+\Psi}{2}$ and $\Phi_{-}=\frac{\Phi-\Psi}{2}$, where $\Phi$ and $\Psi$ are the metric perturbations, are only reliable at scales $k\geq0.01h$ with $h=H_0/(100$ Km s$^{-1}$ Mpc $^{-1}$) \cite{DGPnumericsolutionscardoso}. The numerical solution is consistent with the dynamical scaling (DS) solution both in the self-accelerating and normal branches, but differs in the asymptotic de Sitter phase of the normal branch where the scaling solution cannot be applied \cite{DGPnumericsolutionscardoso}.\\
 In this work we focus on the matter dominated era where the dynamical scaling solution is in agreement with the numerical solution and  the master variable scales as $\Omega_b\propto a^4$ \cite{nearhorizon}. And with this assumption, we show in Section \ref{sec:scalarperturbations} that using the boundary condition found in \cite{DGPnumericsolutionscardoso}, then we obtain a second-order differential equation only for matter density perturbations. To solve this equation for density perturbations, we first constrain the background parameters for both models. For that purpose, in Section \ref{sec:statistical}, we perform a Bayesian statistical analysis, using the Pantheon+ catalogue \cite{Brout_2022} and a mock catalogue for gravitational waves.  
Once we constrain the background model parameters for the normal and self-accelerating branch, we use the RSD observations to constrain the value of $\sigma_8$ in Section \ref{sec:fsigma8}. \\
In Section \ref{sec:results} we show our results, and finally in Section \ref{sec:conclusions} we write the conclusions.


\section{Scalar perturbations}
\label{sec:scalarperturbations}
The most general action of the model is
\begin{equation}
    S=\frac{1}{2\kappa_5^2}\int_{\mathcal{M}}d^5X\sqrt{-g}(R^{(5)}+\Lambda_5)+\frac{1}{2\kappa_4^2}\int_{\partial \mathcal{M}_b}\sqrt{-\gamma}(R^{(4)}+\mathcal{L}-\sigma),
\end{equation}
where $\sigma$ is the brane tension, $R^{(5)}$, $R^{(4)}$ is the five-dimensional and four-dimensional Ricci scalar respectively, $\kappa_4^2=8\pi G_4$, $G_4$ is the 4-dimensional gravitational constant, $\kappa_5^2=8\pi G_5$ and $G_5$ is the 5-dimensional bulk gravitational constant, and the crossover scale is $r_c=\frac{\kappa_5^2}{2\kappa_4^2}$. And in this work, we consider $\Lambda_5=0$.\\ 
The metric for the background is given by
\begin{equation}
    ds^2=-n(y,t)^2dt^2+b(y,t)^2dx^2+dy^2,
\end{equation}
where
\begin{eqnarray}
\label{nyb}
n(y,t)&=&1+\epsilon\left(\frac{\dot{H}}{H}+H\right)|y|,\nonumber\\
b(y,t)&=&a(1+\epsilon H|y|),
\end{eqnarray}
where $\epsilon=1$ for the accelerated branch and $\epsilon=-1$ for the normal branch, and the dot indicates derivative with respect to $t$. \\
Using the junction conditions across the brane, it can be found that the  modified Friedmann equation \cite{deffayet2001cosmology}: 
\begin{equation}
\label{ec:friedmann}
    H^2-\epsilon\frac{H}{r_c}=\frac{8\pi G}{3}(\rho+\sigma),
\end{equation}
and the continuity equation is satisfied: 
\begin{equation}
    \dot{\rho}+3H(\rho+p)=0.
\end{equation}
If we consider only scalar perturbations, the five-dimensional perturbed metric is \cite{onbraneworldcedric}: 
\begin{equation}
g_{AB}=
-n^2(1+2A)dt^2+b^2(1+2\mathcal{R})dx^2+nA_ydydt+(1+2A_{yy})dy^2,
\end{equation}
where $A$, $\mathcal{R}$, $A_y$, $A_{yy}$ are scalars.\\
All gauge-invariant perturbations in the 5D-dimensional bulk can be described using the master variable $\Omega$ that satisfies the following partial differential equation \cite{Mukohyama}: 
\begin{equation}
\label{mastereq}
    -\left(\frac{1}{nb^3}\dot{\Omega}\right)^{.}+\left(\frac{n}{b^3}\Omega'\right)'-\frac{n}{b^5}k^2\Omega=0,
\end{equation}
where the primes indicate derivative with respect to $y$.\\
On the other hand, the perturbed metric on the brane in the Newtonian gauge is: 
\begin{equation}
\label{perturbedmetricbrane}
    ds_b^2=-(1+2\Psi)dt^2+a^2(1+2\Phi)\delta_{ij}dx^idx^j,
\end{equation}
it can be shown that the effective on-brane equations of motion are given by \cite{maeda2003effective}: 
\begin{equation}
\label{background}
    G_{\mu\nu}^{(4)}=(16\pi Gr_c)^2\Pi_{\mu\nu}-\mathcal{E}_{\mu\nu},
\end{equation}
where $\mathcal{E}_{\mu\nu}$ is the projection of the 5D traceless Weyl tensor onto the brane and $\Pi_{\mu\nu}$ is given by: 
\begin{eqnarray}
\label{equationsbrane}
    \Pi_{\mu\nu}&=&-\frac{1}{4}\tau_{\mu\alpha}\tau^{\alpha}_{\nu}+\frac{1}{12}\tau\tau_{\mu\nu}+\frac{1}{8}g_{\mu\nu}\tau_{\alpha\beta}\tau^{\alpha\beta}-\frac{1}{24}g_{\mu\nu}\tau^2,\nonumber\\
    \tau^{\mu}_{\nu}&=&T^{\mu}_{\nu}-(8\pi G)^{-1}G^{\mu(4)}_{\nu}
\end{eqnarray}
where 
$T^{\mu}_{\nu}$ is the energy-momentum tensor on the brane and $\tau$ is the trace of $\tau^{\mu}_{\nu}$.\\
In the background spacetime $\mathcal{E}_{\mu\nu}=0$, but this doesn't happen in the perturbed spacetime. 
From (\ref{background}) we can obtain the perturbed on-brane equations given by: 
\begin{equation}
\label{perturbedeinstein}
    \delta G^{(4)}_{\mu\nu}=(16\pi Gr_c)^2\delta\Pi_{\mu\nu}-\delta\mathcal{E}_{\mu\nu},
\end{equation}
where $\delta\Pi_{\mu\nu}$ can be obtained using the perturbed metric on the brane given by equation (\ref{perturbedmetricbrane}) and the perturbed energy-momentum tensor for a fluid given by \cite{onbraneworldcedric}: 
\begin{equation}
    \delta T^{\mu}_{\nu}=\begin{pmatrix}
-\delta \rho & a \delta q_{,i}\\
-a^{-1}\delta q^{,i} & \delta p\delta ^{i}_j\\
\end{pmatrix}.
\end{equation}
While the perturbations of the Weyl tensor can be considered as perturbations of an effective fluid, as: 
\begin{equation}
    \delta \mathcal{E}^{\mu}_{\nu}=-8\pi G\begin{pmatrix}
-\delta \rho_{\mathcal{E}} & a \delta q_{\mathcal{E},i}\\
a^{-1}\delta q_{\mathcal{E}}^{,i} & \frac{1}{3}\delta\rho_{\mathcal{E}}\delta ^{i}_j+\delta\pi^{i}_{\mathcal{E}j}\\
\end{pmatrix},
\end{equation}
and it can be shown that the Weyl fluid perturbations are related to $\Omega$ through
\begin{eqnarray}
\label{weylmaster}
    \kappa_4^2\delta\rho_{\varepsilon}=\frac{k^4\Omega_b}{3a^5}, \hspace{0.5cm}\kappa_4^2 a\delta q_{\varepsilon}=-\frac{k^2}{3a^3}(H\Omega_b-\dot{\Omega_b}),
\end{eqnarray}
where $\Omega_b$ is the value of $\Omega$ on the brane, that is to say at $y=0$.\\
Then from the $(0,0)$ component of equation (\ref{perturbedeinstein}), it can be found the modified Poisson equation:
\begin{equation}
\label{modifiedpoisson}
    \frac{k^2}{a^2}\Phi=4\pi G\left(\frac{2\epsilon Hr_c}{2H\epsilon r_c-1}\right)\left(\rho\triangle-\frac{\delta\rho_{\mathcal{E}}-3aH\delta q_{\mathcal{E}}}{2\epsilon Hr_c}\right),
\end{equation}
where $\rho\triangle=\delta\rho-3aH\delta q$
while from the $(0,i)$ component of equation  (\ref{perturbedeinstein}) it can be shown that
\begin{equation}
    H\Psi-\dot{\Phi}=\frac{4\pi G}{2Hr_c\epsilon-1}(\delta q_{\mathcal{E}}-2\epsilon Hr_c\delta q).
\end{equation}
The Poisson equation can be used to obtain a boundary condition for $\Omega$ given by \cite{DGPnumericsolutionscardoso}: 
\begin{eqnarray}
\label{boundaryomega}
    (\partial_y\Omega)_b=-\frac{\epsilon\gamma_1}{2H}\ddot{\Omega}_b+\frac{9\epsilon\gamma_3}{4}\dot{\Omega}_b-\left(\frac{3\epsilon\gamma_3k^2}{4Ha^2}+\frac{H\gamma_4}{4}\right)\Omega_b+\frac{3\epsilon r_c\kappa_4^2\rho a^3\gamma_4}{2k^2}\triangle,
\end{eqnarray}
where $\gamma_1$,$\gamma_3$ and $\gamma_4$ are defined in the Appendix \ref{app:scalingsolution}  in equation (\ref{ec:gammas}).\\
From the modified Poisson equation (\ref{modifiedpoisson}) and using (\ref{weylmaster}), we can find that  $\Phi$ in terms of $\Omega_b$ is: 
\begin{equation}
\label{phimaster}
    \Phi=\frac{\kappa_4^2\rho a^2\gamma_1\triangle}{2k^2}+\frac{\epsilon\gamma_1}{4ar_c}\dot{\Omega_b}-\epsilon\left(\frac{k^2}{12Hr_ca^3}+ \frac{H}{4ar_c} \right)\gamma_1\Omega_b
    \end{equation}
    and $\Psi$ can be obtained using equations (\ref{Psi}) and (\ref{boundaryomega}), then
    \begin{equation}
    \label{psimaster}
       \Psi=-\frac{\kappa_4^2\rho a^2\gamma_2}{2k^2}\triangle+\frac{\epsilon\gamma_1}{4Hr_ca}\ddot{\Omega}_b-\frac{3\epsilon H\gamma_4}{4a}\dot{\Omega}_b+\epsilon\left(\frac{k^2\gamma_4}{4a^3}+\frac{H\gamma_2}{4ar_c}\right)\Omega_b.
\end{equation}
On the other hand, from the conservation equations: 
\begin{equation}
    \delta(\nabla^{\alpha}T_{\alpha\beta})=0,
\end{equation}
it can be obtained: 
\begin{eqnarray}
\label{dotq}
    \delta\dot{q}=-4H\delta q-\frac{\delta p}{a}-\frac{\Psi}{a}(\rho+p)\nonumber\\
    \frac{\delta\dot{\rho}}{\rho}=-\frac{\nabla^2\delta q}{a\rho}-3\dot{\Phi}(1+\omega)-3H\left(\frac{\delta\rho}{\rho}+\frac{\delta p}{\rho}\right).
   \end{eqnarray} 
   If we consider on the brane, only dark matter and if we combine the equations (\ref{dotq}), and we define $\rho_m\triangle_m={\delta\rho_m}-3aH\delta q_m$ where $\rho_m$ is the dark matter density. We can find a second-order differential equation for $\triangle_m$ \cite{DGPnumericsolutionscardoso}:
   \begin{equation}      
   \label{triangledifeq}
   \ddot{\triangle}_m+2H\dot{\triangle}_m=-\frac{k^2}{a^2}\Psi+\frac{3}{2}\dot{F}+3HF,
   \end{equation}
   where 
   \begin{equation}
    F=\frac{\kappa_4^2a\delta q_{\varepsilon}}{2Hr_c\epsilon-1}.
   \end{equation}
Then replacing (\ref{psimaster}) and (\ref{weylmaster}), in (\ref{triangledifeq}) we obtain:
\begin{equation}
\label{tri}
    \ddot{\triangle}_m+2H\dot{\triangle}_m-\frac{1}{2}\kappa_4^2\rho_m\gamma_2\triangle_m=-\epsilon\frac{\gamma_4k^4}{4a^5}\Omega_b,
\end{equation}
As usual, we define the density parameters: 
\begin{eqnarray}
\label{densityparameters}
    \Omega_{m}=\frac{\rho_m}{\rho_c}=\frac{\Omega_{m0}}{a^3}\left(\frac{H_0}{H}\right)^2,\hspace{0.5cm}\Omega_{r0}=\frac{1}{4r_c^2H_0^2},
\end{eqnarray}
where $\rho_c$ is the critical density and $\rho_c=(8\pi G_4)/(3H^2)=\kappa_4^2/(3H^2)$, $\Omega_m$ is the density parameter of dark matter and $\Omega_{m0}$ its corresponding present value.
If we replace (\ref{densityparameters}) in  (\ref{tri}), then we obtain:
\begin{equation}
\label{eqtrianglea}
\frac{d^2\triangle_m}{da^2}+\left(\frac{3}{a}+\frac{1}{H}\frac{dH}{da}\right)\frac{d\triangle_m}{da}=\frac{3}{2}\frac{H_0^2\Omega_{m0}\gamma_2\triangle_m}{a^5H^2}-\epsilon\frac{\gamma_4k^4\Omega_b}{4H^2a^7},
\end{equation}
where $\gamma_2$ and $\gamma_4$ are given in the Appendix \ref{app:scalingsolution} in terms of $\Omega_{r0}$ and $a$. 
From the above equation, it can be seen that once we know $\Omega_b$, we can solve (\ref{eqtrianglea}). But to obtain $\Omega_b$ we have to solve (\ref{mastereq}) with boundary condition (\ref{boundaryomega}). As we have already mentioned in the introduction, in the literature, there are different ways to solve it and in this work to solve it we assume the scaling solution $\Omega=\mathcal{A}a^p G$, see Appendix \ref{app:scalingsolution}, with $p=4$. When we replace $\Omega=\mathcal{A}a^pG(x)$ with $x=yH$ in (\ref{mastereq}) we obtain a second differential equation for $G$ that can be solved numerically as a boundary value problem from $x=0$ to $x=1$ and with boundary conditions $G(x=0)=1$ and $G(x=1)=0$.

The second-order differential equation for $G$ is given by \cite{largescalesong} 
\begin{equation}
\label{G}
    A(x)\frac{d^2G}{dx^2}+B(x)\frac{dG}{dx}+C(x)G=0,
\end{equation}
where 
\begin{eqnarray}
\label{coefficientnormaldG}
A(x)&=&(1-x)(1-x-2hx),\nonumber\\
B(x)&=&-2x(hp+1)+2-h+\frac{(x^2-x)(h^2+\tilde{h}+h)}{1-x(h+1)},\nonumber\\
C(x)&=&-p^2-hp-\frac{xp(\tilde{h}+h^2+h)}{1-x(h+1)}+\frac{3p(1-x-xh)}{1-x}-\frac{[1-x(1+h)]^2}{(1-x)^2}\frac{k^2}{a^2H^2},
\end{eqnarray}
for the normal branch. Here $h=(dH/d\ln a)/H$ and $\tilde{h}=dh/(d\ln a)$.\\
While for the accelerated branch:
\begin{eqnarray}
\label{coefficeintaccdG}
    A(x)&=&(1+x)(1+x(1+2h))\nonumber\\
    B(x)&=&-2x(hp+1)-2+h-\frac{(x^2+x)(h^2+\tilde{h}+h)}{1+x(h+1)}\nonumber\\
    C(x)&=&-p^2-hp+\frac{xp(\tilde{h}+h^2+h)}{1+x(h+1)}+\frac{3p(1+x+xh)}{1+x}-\frac{[1+x(h+1)]^2}{(1+x)^2}\frac{k^2}{a^2H^2}.
\end{eqnarray}
Furthermore, if we replace $\Omega=\mathcal{A}a^p G(x)$ in the equation for the boundary condition (\ref{boundaryomega}) and using (\ref{ec:dotomega}), we can find that: 
\begin{equation}
\label{ec:dGdx1}
    \frac{dG}{dx}|_{y=0}=-\frac{\epsilon\gamma_1}{2}(h^2+hp)+\frac{9\epsilon\gamma_3}{4}p-\frac{3\epsilon\gamma_3k^2}{4H^2a^2}-\frac{H\gamma_4}{4}+\frac{3\epsilon r_c\kappa_4^2\rho a^3\gamma_4}{2k^2}\frac{\triangle}{\mathcal{A}a^p H},
\end{equation}
if we consider $\rho=\rho_m$, then (\ref{ec:dGdx1}) can be rewritten in terms of the density parameters as: 
\begin{equation}
\label{ec:dGdx2}
    \frac{dG}{dx}|_{y=0}=-\frac{\epsilon\gamma_1}{2}(h^2+hp)+\frac{9\epsilon\gamma_3}{4}p-\frac{3\epsilon\gamma_3k^2}{4H^2a^2}-\frac{H\gamma_4}{4}+\frac{9H_0\Omega_{m0}}{4\sqrt{\Omega_{r0}}k^2}\frac{\gamma_4\triangle_m}{\mathcal{A}a^p H},
\end{equation}
where $\gamma_1,\gamma_3$ and $\gamma_4$ are given in terms of $a$ and $\Omega_{r0}$ in Appendix \ref{app:scalingsolution}.
Hence, once we find numerically $G$, we can compute numerically $(dG/dx)$ and therefore $(dG/dx)|_{y=0}$ and substitute this value in (\ref{ec:dGdx2}) to obtain $\Omega_b=\mathcal{A}a^p$ from: 
\begin{equation} 
\label{omegasol}
\Omega_b=\frac{\delta\triangle_m\gamma_4\epsilon}{H\left(\frac{dG}{dx}|_{y=0}+\frac{\epsilon\gamma_1(h^2+hp)}{2}-\frac{9\epsilon\gamma_3p}{4}+\frac{3\epsilon\gamma_3 k^2}{4a^2H^2}+\frac{3H\gamma_4}{4}\right)},
\end{equation}
where 
\begin{equation}
    \delta=\frac{9 H_0\Omega_{m0}}{4\sqrt{\Omega_{r0}}k^2 }.
\end{equation} 
Then, if we replace expression (\ref{omegasol}) in equation (\ref{eqtrianglea}), we find a second-order differential equation only for $\triangle_m$, which we can solve numerically.
Therefore, we have found that if we use the second differential equation for matter density perturbations (\ref{eqtrianglea}), the boundary condition found previously in \cite{DGPnumericsolutionscardoso} and at the same time we use the scaling solution, then we obtain a second order differential equation only for $\Delta_m$ that we can solve numerically. 
 And finally once we know $\triangle_m=\frac{\delta\rho_m}{\rho_m}-3aH\frac{\delta q_m}{\rho_m}$ we can obtain $\Phi$ and $\Psi$ from (\ref{phimaster}), and the growth rate $f$ defined in  subsection \ref{sec:fsigma8}. \\
To solve (\ref{eqtrianglea}), we first constrain the background parameters using the Supernovae and gravitational wave observations for the normal and the self-accelerating branches.
For that purpose, we perform a Bayesian statistical analysis to obtain the best-fit parameter values of the models, which is described in Section \ref{sec:statistical}.\section{Statistical analysis and data}
\label{sec:statistical}
To obtain the value of the background parameters for the normal and self-accelerating branch, we perform a statistical Bayesian analysis using the following catalogues, which we describe briefly below:
\subsection{Pantheon +}

 In this sample are presented the 1701 light curves of 1550 Type Ia Supernovae (SNe Ia) in a redshift range $z\epsilon[0.0001,2.26]$ from 18 different surveys \cite{Scolnic_2022}. In our analysis, we use the data collected by \cite{brout2022pantheon+}, this data is available at this URL: https://github.com/PantheonPlusSH0ES/DataRelease/tree/main/Pantheon. Also, the covariance matrix $C_{stat+syst}$ is included on this page, which includes the statistical and systematic uncertainties.\\
Data includes the apparent magnitude in the B band $m_{B}$ of the Supernovae as well as their uncertainty.

The theoretical distance modulus $\mu$ and $m_B$ are related by: 
\begin{equation}
    \mu(z)=m_B-M,
\end{equation}
where $M$ is the SnIa absolute magnitude. On the other hand $\mu$ is related to the luminosity distance, $d_L$ as follows: 
\begin{equation}
    \mu(z)=5 log\left[\frac{d_L(z)}{1 Mpc}\right]+25,
\end{equation}
and $d_L$ is given by the following expression: 
\begin{equation}
\label{dL}
    d_L=a_0c(1+z)\int_{0}^z\frac{dz}{H},
\end{equation}
where $H$ is given by (\ref{Hnormal}) for the normal branch and by (\ref{Hself}) for the self-accelerating branch.\\
As data also includes the distance modulus of the Cepheid hosts $\mu_i^{Ceph}$ of the $i^{th}$ SnIa, which is measured independently by the SH0ES team \cite{Riess2021ACM}, then
$\mu_i^{Ceph}=m_{Bi}-M$. 
Then the best-fit parameters for a specific model, using the Pantheon+ catalogue,
can be calculated by maximizing the logarithm of the likelihood function, or equivalently by minimizing the $\chi^2$ likelihood given by: 
\begin{equation}
\chi^2=\vec{Q}^T\cdot(C_{stat+syst})\cdot \vec{Q},
\end{equation}
where $\vec{Q}$ is a vector of dimension $1701$ and whose components are defined as:
\begin{equation}
Q_i = \left\{
	       \begin{array}{ll}
		 m_{Bi}-M-\mu_{i}^{Ceph}      & i\hspace{0.1cm}\epsilon\hspace{0.1cm}\mathrm{Cepheid}\hspace{0.1cm}\mathrm{hosts} \\
		 m_{Bi}-M-\mu(z_i)& \mathrm{otherwise}.
	       \end{array}\right.
\end{equation}

\subsection{Gravitational waves mock data}

We use data from a mock catalogue, which
consists of standard sirens mock data based on the Laser Interferometer Space Antenna (LISA) by forecasting multimessenger measurements of massive black hole binary (MBHB) mergers \cite{Corman:2021avn,Corman:2020pyr}. This catalogue includes the gravitational wave luminosity distance denoted by $d_L^{GW}$ of 1000 simulated events with their respective redshifts and errors.

And we can compute the best-fit parameters of a model, minimizing the $\chi^2$ likelihood function: 
\begin{equation}
\label{chiGW}
    \chi^2_{\text{GW}}= \sum_{i=1}^{1000}\frac{(d_L^{\text{GW}}(z_i,\Theta)-d_{Lm}^{\text{GW}}(z_i))^2}{\sigma_{i}^2},
\end{equation}
where $d_L^{GW}(z_i)$ is the theoretical gravitational wave luminosity distance of the model at redshift $z_i$ and $d_{Lm}^{GW}(z_i)$ is the gravitational wave luminosity distance obtained from the mock catalogue at redshift $z_i$ and  $\sigma_{i}$ is its corresponding error. And $\Theta$ are the free parameters of the model. \\ 
 As it is shown in \cite{hernandezmarquez}, since  within the DGP framework the 4-dimensional brane is embedded in a 5-dimensional Minkowski space-time and gravity can propagate through this extra dimension, the gravitational wave luminosity distance, $d_L^{GW}$, the distance measured from gravitational events, e.g. binary BH coalescence is different from the electromagnetic luminosity
distance $d_L$ through: 
\begin{equation}
    d_L^{GW}=d_L\left[1+\left(2H_0\sqrt{\Omega_{r0}}\frac{d_L}{c(1+z)}\right)^m\right]^\frac{1}{2m},
\end{equation}
where $d_L$ is given by (\ref{dL}) and $m$ determines the steepness of the transition from the small-scale to large-scale behavior and is a free parameter that has to be determined \cite{Corman:2021avn}.
\subsection{\texorpdfstring{$f\sigma_8$}{f\sigma_8}}
\label{sec:fsigma8}
The growth rate $f$ is defined as 
\begin{equation}
    f(a)\equiv\frac{d\ln\triangle_m(a)}{d\ln a}.
\end{equation}
However, in the past two decades, the vast majority of LSS surveys report instead the bias-independent product $f\sigma_8(a)=f(a)\cdot\sigma_8(a)$, where 
\begin{equation}
    \sigma_8(a)\equiv\frac{\sigma_8}{\triangle_m(1)}\triangle_m(a),
\end{equation}
with $\sigma_8$ corresponding to the density root mean square (\text{rms}) fluctuations within spheres on scales of about $8h^{-1}$Mpc, formally $\sigma_8$ is defined as: 
\begin{equation}
{\sigma_8}^2 = \int^{\infty}_{0} k^2 P(k)W^2(kR_8)dk,
\end{equation}
where $R_8 =8h^{-1}$Mpc, $P(k)$ is the linear matter power spectrum today and $W(kR_8)$ is the Fourier transform of the top-hat window \cite{huterer2023course}, and $\triangle_m$ is the solution to the differential equation (\ref{eqtrianglea}) with $\Omega_b$ given by (\ref{omegasol}).\\
Then 
\begin{equation}
    \label{fsigma8a}f\sigma_8(a)=a\frac{d\triangle_m(a)}{da}\frac{\sigma_8}{\triangle_m(1)}.
\end{equation}
From (\ref{fsigma8a}) we can see that the only free parameter is $\sigma_8$, which we want to determine for the normal and the self-accelerating branch. For that purpose, we perform a Bayesian statistical analysis using the data of $f\sigma_8(a)$ presented in Table \ref{tab:rsd}, these entries  were selected from a larger observational data compilation reported in \cite{alestas2022machine} based on criteria of data quality, independence and relevance of the redshift range and scales of interest for constraining the growth rate $f\sigma_8$. The chosen 15 points cover a representative range of redshifts from very low ($z\approx0.013)$ to moderately high $(z\approx1.48)$, allowing effective constraints on cosmological models over time while avoiding highly non-linear scales or data with large uncertainties which could dilute the constraints. In essence, the selected data are a curated subset optimized for the combined goals of maximizing independent constraining power and minimizing systematic issues as recommended in prior growth rate analyses \cite{clusteringSDSS}, \cite{blakeetal2011}.  
To determine $\sigma_8$, we maximize the logarithm of the likelihood function given by: 
\begin{equation}
    \ln \mathcal{L}_{f\sigma_8}(f\sigma_8(a_i)|a_i,\sigma_i,\sigma_8)=-\frac{1}{2}\left(\chi_{f\sigma_8}^2+\sum_{n=1}^{15}\ln(2\pi\sigma_i)^2\right),
    \label{eq:likelihoodfsigma8}
\end{equation}
where 
\begin{equation}   \chi_{f\sigma_8}^2=\sum_i^{15}\frac{f\sigma_8(a_i,\sigma_8)-f\sigma_8^{obs}(a_i)}{\sigma_i^2},
    \label{eq:chi2fsigma8}
\end{equation}
where $\sigma_i$ is the variance of each measurement.\\
To constrain $\sigma_8$ according to (\ref{eq:chi2fsigma8}) we need to compute (\ref{fsigma8a}) for each $a_i$ of Table \ref{tab:rsd} and for that we have to solve (\ref{eqtrianglea}) as it is described in Section \ref{sec:scalarperturbations}. We do this using the best-fit values for the parameters of the models shown in Table \ref{tab:nor1} and Table \ref{tab:nor2}  for the normal branch and in Table \ref{tab:self_acc} for the self-accelerating branch. We set as initial condition $\triangle_m/a_i=1$, where $a_i$ is the initial value of $a$ and because we are interested in the matter-dominated era, $a_i=0.01$. As $\sigma_8$ corresponds to the density rms fluctuations with spheres of radius on scales of about $8h^{-1}$Mpc, we solve this differential equation for $k=(h/8)$Mpc$^{-1}$ where $h=H_0/(100$ km s$^{-1}$ Mpc$^{-1}$)  and $H_0$ is the best-fit value found for the sum of data shown in Table \ref{tab:nor1} and Table \ref{tab:nor2} for the normal branch, and Table \ref{tab:self_acc} for the self-accelerating branch.\\
It is important to mention that the substitution of $k=(h/8)$ Mpc$^{-1}$ in equation \ref{eqtrianglea} samples $\triangle_m$ at the wavelength corresponding to this scale, as it can be seen equation \ref{eqtrianglea} of $\triangle_m$  depends on the scale and therefore $f\sigma_8$ (eq. \ref{fsigma8a}) depends on $k$. This approximation is commonly employed in cosmological literature when the analysis centers on the relative growth of perturbations on cluster scales, especially if the matter power spectrum shape is close
to standard forms and the focus is on the growth factor evolution rather than a full power spectrum integration. For instance, \cite{ishak2006probing} employ a similar approach in testing modified gravity models, using the growth factor evaluated at a characteristic scale to approximate $\sigma_8$ variations, arguing it captures the main behavior relevant for comparing model predictions with observations. The validity of this approximation relies on the scale dependence of growth being mild around $k=h/8$ Mpc$^{-1}$, but as it can be seen in Figure \ref{fig:growthfactor} the evolution of $\triangle_m$ is the same for different values of $k$ in both branches. And this is confirmed in Figure \ref{fig:difference} where we plot the difference of $\triangle_m/a$ between $k=0.125h$ and $k=0.002h$ and it can be seen that is approximately zero.
 \begin{table}
\centering
\begin{tabular}{|c|c|c|}
\hline
     $z$&$a$&$f\sigma_8^{obs}$  \\
     \hline
     $0.013$&$0.987$&$0.46\pm0.06$\\
     \hline
     $0.02$&$0.980$&$0.428\pm0.048$\\
     \hline
     $0.15$&$0.869$&$0.53\pm0.16$\\
     \hline
     $0.17$&$0.854$&$0.51\pm0.06$\\
     \hline
     $0.18$&$0.847$&$0.36\pm0.9$\\
     \hline
     $0.38$&$0.725$&$0.5\pm0.047$\\
     \hline
     $0.44$&$0.694$&$0.413\pm0.08$\\
     \hline
     $0.51$&$0.662$&$0.455\pm0.39$\\
     \hline
     $0.6$&$0.625$&$0.55\pm0.12$\\
     \hline
     $0.7$&$0.588$&$0.448\pm0.043$\\
     \hline
     $0.73$&$0.578$&$0.437\pm0.072$\\
     \hline
     $0.85$&$0.540$&$0.315\pm.095$\\
     \hline
     $0.86$&$0.537$&$0.4\pm0.11$\\
     \hline
     $1.4$&$0.416$&$0.482\pm0.116$\\
     \hline
     $1.48$&$0.403$&$0.462\pm .045$\\
     \hline
\end{tabular}
\caption{Observational values for $f\sigma_8$ obtained from RSD observations and compiled in \cite{alestas2022machine}.} 
\label{tab:rsd}
\end{table}


\section{Analysis and results}
\label{sec:results}
To constrain the background parameters and compute their respective posterior distributions of the normal and self-accelerating branch, we perform a Bayesian statistical analysis with the Pantheon+ catalogue, the mock catalogue of standard sirens described in Section \ref{sec:statistical} and the combination of both catalogues. To perform the analysis, we use the \texttt{emcee}
\footnote{\href{https://emcee.readthedocs.io/en/stable/}
{emcee.readthedocs.io}} code and we combine the marginalised distributions for each fractional density of the models using the \texttt{ChainConsumer} \footnote{\href{https://samreay.github.io/ChainConsumer/}{samreay.github.io/ChainConsumer}} package.


\subsection{\texorpdfstring{Normal branch: $\Lambda$ DGP}{Normal branch: \Lambda DGP}}
\begin{figure}
    \centering
    \includegraphics[width=12cm]{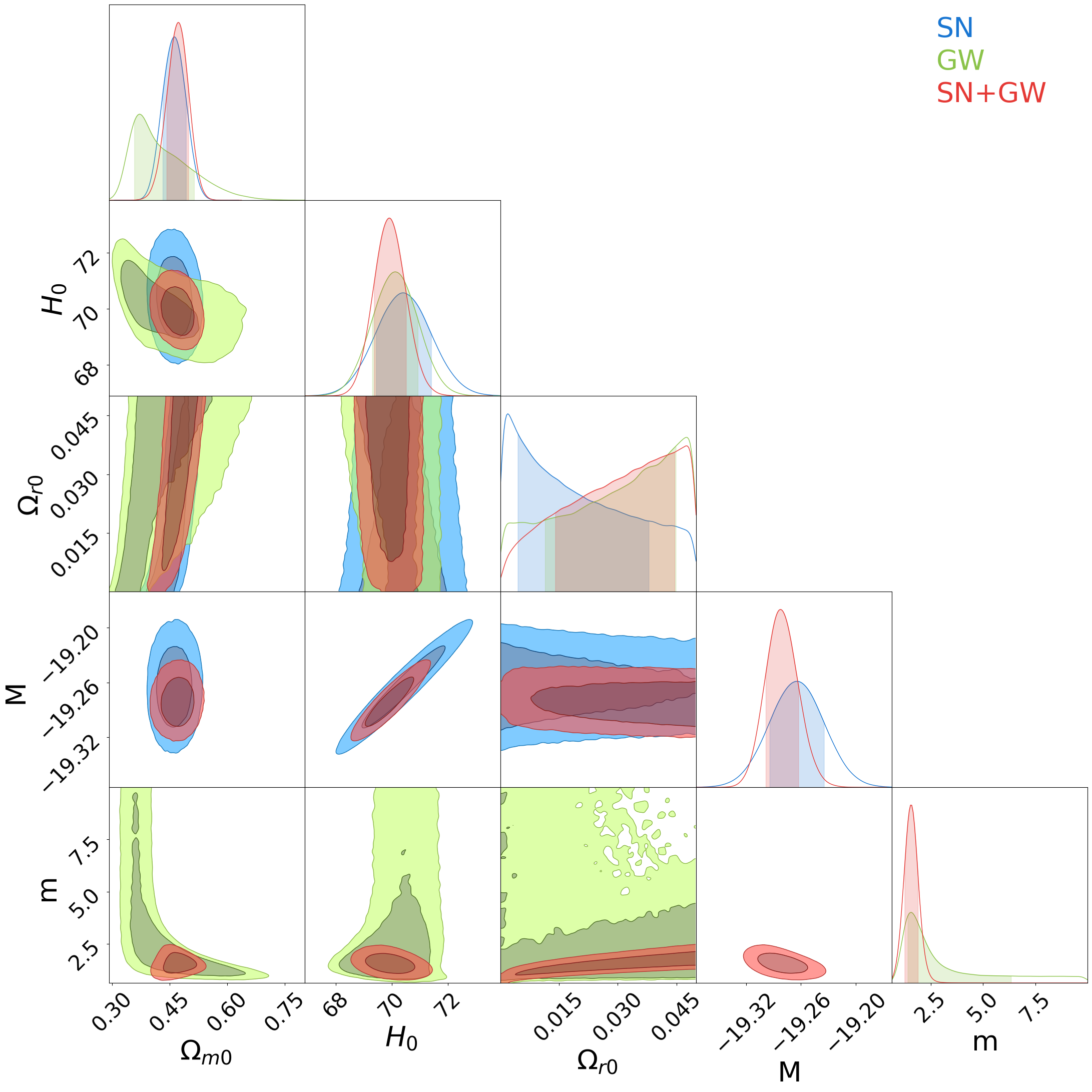}
    \caption{$2\sigma$ C.L. constraints for the background parameters using standard sirens mock data GW (green) and SN Pantheon+ (blue), and the total sample GW+Pantheon+ (red) for the normal branch using the priors shown in Table \ref{tab:nor2}.}
    \label{fig:posteriors_normal}
\end{figure}

For this model, the Friedmann equation is obtained by replacing $\epsilon=-1$ and considering the tension $\sigma\neq0$ (\ref{ec:friedmann}), with this, the tension acts as an effective $4-$dimensional cosmological constant, and there is a late time accelerating phase.
Then the Friedmann equation in terms of the density parameters can be written as: 
\begin{equation}
\label{Hnormal}
    H=H_0\left(\sqrt{\frac{\Omega_{m0}}{a^3}+\Omega_{\sigma}+\Omega_{r0}}-\sqrt{\Omega_{r0}}\right)
\end{equation}
where 
$\Omega_{\sigma}=\frac{\kappa_4^2\sigma}{3H_0^2}=1-\Omega_{m0}+2\Omega_{r0}^{1/2}$, $H_0$ is the current value of Hubble constant. According to Section \ref{sec:statistical} the background cosmological parameters for this model with data of Pantheon+ are: $\Omega_{m0}$, $H_0$, $\Omega_{r0}$ and $M$, while for the mock catalog of gravitational waves are $\Omega_{m0}$, $\Omega_{r0}$, $H_0$ and $m$.  Additionally, according to theory $r_c\backsim H_0^{-1}$, then $\Omega_{r0}=\frac{1}{4r_c^2H_0^2}\backsim0.25$ but if the crossover scale were larger $r_c\gtrsim H_0^{-1} $then $\Omega_{r0}\lesssim0.25$. So in  order to constrain the value of $\Omega_{r0}$ we first assume a uniform prior such that $\Omega_{r0}\epsilon(10^{-6},0.25$).\\
In Table \ref{tab:nor1} are shown the best-fit values for the background parameters obtained from the statistical analysis for the Pantheon+ catalogue labelled as SN, the mock catalogue of gravitational waves labelled as GW and the sum of data labelled as SN+GW. We found that using this prior for $\Omega_{r0}$ then the best-fit values for the background parameters are $\Omega_{m0}=0.546\pm0.039$, $H_0=(69.58\pm0.067)$ Km s$^{-1}$ Mpc$^{-1}$ and $\Omega_{r0}=0.155\pm0.067$, $M=-19.293\pm0.016$ and $m=3.7\pm 1.5$, for the sum of data. However, in the $\Lambda$CDM model with flat spatial curvature $\Omega_k=0$ using data from Pantheon+ and SH0ES $\Omega_{m0}=0.334\pm0.018$ and $H_0=(73.6\pm1.1)$ Km s$^{-1}$ Mpc$^{-1}$ \cite{brout2022pantheon+}, then this value of $\Omega_{m0}$ is greater than the one obtained in the $\Lambda$CDM model.  \\
On the other hand in \cite{lombriser2009cosmological} using data of CMB it was found that $\Omega_{r0}<0.05$, then we use a second uniform prior such that $\Omega_{r0}\epsilon(0,0.05)$ and we found that for the sum of data $\Omega_{m0}=0.470\pm0.028$,  $H_0=(69.94\pm0.57)$Km s$^{-1}$ Mpc$^{-1}$, $\Omega_{r0}=0.029\pm0.015$, $M=-19.281\pm0.018$ and $m=1.57\pm0.32$ (see  Table \ref{tab:nor2}). Therefore, in this branch, the value of $\Omega_{r0}$ has to be small to have a lower matter density parameter; however, $\Omega_{m0}$ is still greater than in the $\Lambda$CDM model. The posteriors for the background parameters using the priors of Table \ref{tab:nor2} are shown in Figure \ref{fig:posteriors_normal}.\\
With the best-fit values shown in Table \ref{tab:nor1} and Table \ref{tab:nor2} for the sum of data, we solve (\ref{eqtrianglea}) with $p=4$ and using (\ref{ec:dGdx2}) and (\ref{omegasol}). With this, we obtain $\triangle_m$ and in Figure \ref{fig:growthfactor} (left side) we show the evolution of the growth factor $\triangle_m/a$ for different values of $k$. We can see from this Figure that the growth factor is affected by the current amount of dark matter $\Omega_{m0}$ and by $H_0$ when there is more dark matter and $H_0$ is lower the deviation of the evolution of growth factor is greater than when there is less dark matter and $H_0$ is greater.  On the other hand, it is remarkable to note that the evolution of the growth factor is the same as that found in \cite{DGPnumericsolutionscardoso}, where the complete numerical solution without assuming the scaling ansatz is considered. However, the evolution of the growth factor is different from the $\Lambda$CDM model.

\begin{figure}
    \centering
    \includegraphics[width=5.2cm]{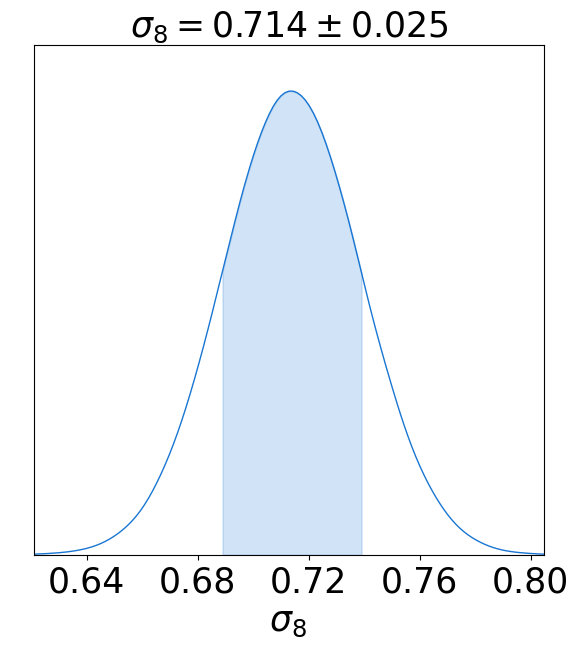}
    \includegraphics[width=5cm]{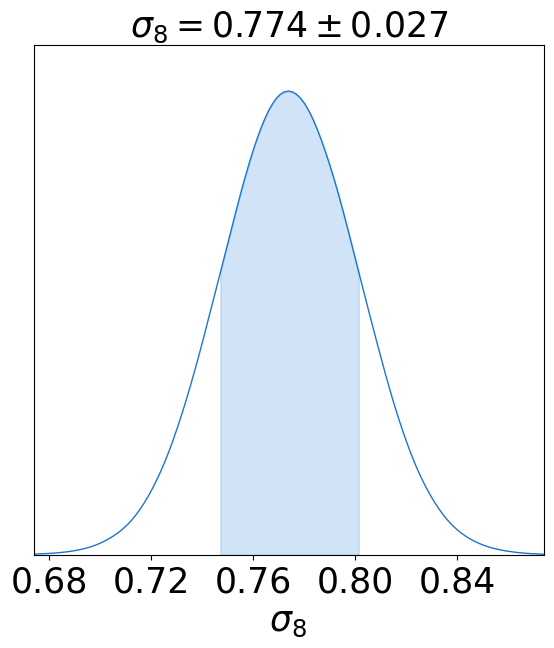}   
    \caption{Left side: Posterior of $\sigma_8$ for the normal branch using data of Pantheon + and a flat prior for $0<\Omega_{r0}<0.25$. Right side: Posterior of $\sigma_8$ for the normal branch using a flat prior for $0<\Omega_{r0}<0.05$. We use a uniform prior for $\sigma_8\epsilon(0,1)$.}
    \label{fig:sigma8normal}
\end{figure}

To constrain the value of $\sigma_8$ we solve numerically the equation for matter density perturbations (\ref{eqtrianglea})  using (\ref{ec:dGdx2}) and (\ref{omegasol}) for $k=(h/8)$Mpc$^{-1}$ with initial condition $\triangle_m/a_i=1$, where $a_i=0.01$. 
Then with $\triangle_m$ we can obtain (\ref{fsigma8a}) and perform the Bayesian statistical analysis using (\ref{eq:likelihoodfsigma8}) and (\ref{eq:chi2fsigma8}).
We obtain the value of $\sigma_8=0.714\pm0.025$ when $\Omega_{m0}=0.546\pm0.039$ and $H_0=(69.58\pm0.067)$ Km s$^{-1}$ Mpc$^{-1}$ and $\sigma_8=0.774\pm0.027$ when $\Omega_{m0}=0.470\pm0.028$  and $H_0=(69.94\pm0.57)$ Km s$^{-1}$ Mpc$^{-1}$. The posteriors of $\sigma_8$ are shown in Figure \ref{fig:sigma8normal}.


\begin{table}
    \centering
    \begin{tabular}{|c|c|c|c|c|c|}
        \hline
		Parameter & Prior &SN&GW&SN+GW\\
  \hline
  $\Omega_{m0}$ &(0, 1)&$0.503\pm 0.054$& $0.62\pm 0.11$ &$0.546\pm 0.039$\\
  \hline
  $\Omega_{r0}$ & (0, 0.25)& $0.087\pm 0.075$&$0.186\pm 0.048$ &$0.155\pm 0.067$\\
  \hline
  $H_0$[Km s$^{-1}$Mpc$^{-1}$] &$(66, 74)$& $70.45\pm 0.98$ &$69.08\pm 0.88$ &$69.58\pm 0.53$\\
  \hline
  M & $(-21, -18)$&$-19.264\pm 0.030$ & --&$-19.293\pm 0.016$\\
  \hline
  m&$(0.1, 10)$&--&  $4.2\pm 2.1$&$3.7\pm 1.5$ \\ 
		\hline
    \end{tabular}
\caption{Best-fit values for the cosmological parameters of the normal branch using the data of Pantheon+ and a flat prior for $0<\Omega_{r0}<0.25$.}
  \label{tab:nor1}
\end{table}

\begin{table}
    \centering
    \begin{tabular}{|c|c|c|c|c|}
        \hline
		Parameters & Prior&SN&GW&SN+GW\\
  \hline
  $\Omega_{m0}$ &$(0,1)$&$0.462\pm 0.030$&$0.435\pm0.078$&$0.470\pm 0.028$\\
  \hline
  $H_0$[Km s$^{-1}$ Mpc$^{-1}$]&(66,74)&$70.42\pm0.98$&$70.12
  \pm 0.81$ &$69.94\pm 0.57$\\
  \hline
   $\Omega_{r0}$&$(0,0.05)$&$0.021\pm 0.017$ & $0.028\pm 0.017$&$0.029\pm0.015$\\
   \hline
   M &$(-21,-18)$&$-19.264\pm 0.030$ &$--$&$-19.281\pm 0.018$\\
   \hline
  m &$(0.1,10)$ &$--$ &$3.9\pm2.5$ &$1.57\pm 0.32$ \\
		\hline
    \end{tabular}
    \caption{Results for the normal branch with the Pantheon+ catalog and using a uniform prior for 
    $0<\Omega_{r0}<0.05$}
    \label{tab:nor2}
\end{table}


\subsection{Self-accelerating branch}
\begin{figure}
    \centering
    \includegraphics[width=8.5cm]{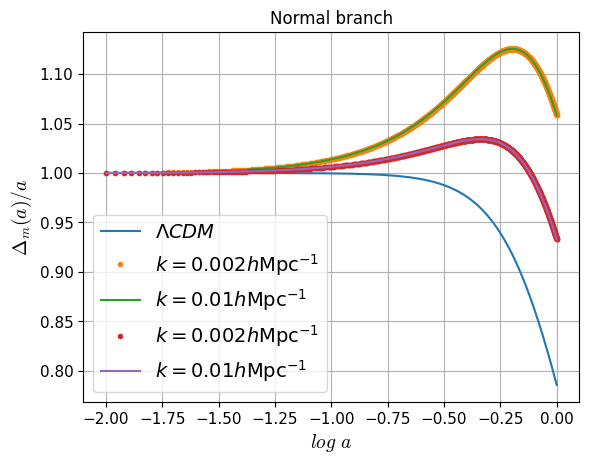}\includegraphics[width=8.5cm]{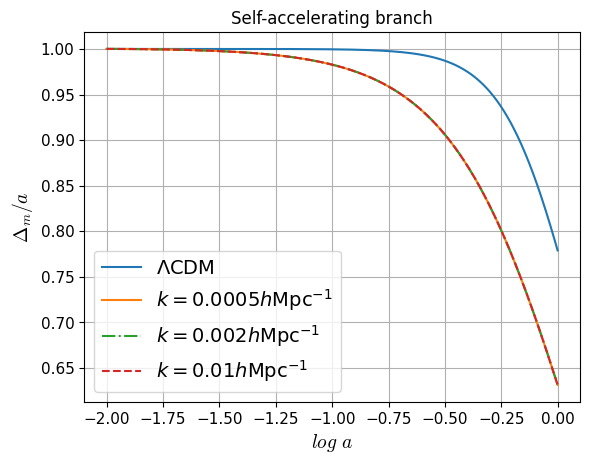}
    \caption{Left side: a) Evolution of $\triangle_m/a$ for the normal branch using the values obtained for the background parameters shown in Table \ref{tab:nor1}: $\Omega_{m0}=0.546$, $H_0=69.58$ Km s$^{-1}$ Mpc$^{-1}$ and $\Omega_{r0}=0.155$ for $k=0.002h$Mpc$^{-1}$ (orange) and $k=0.01h$Mpc$^{-1}$ (green). b) Difference of branch using the values obtained for the background parameters shown in Table \ref{tab:nor2}: $\Omega_{m0}=0.470$, $H_0=69.94$ Km s$^{-1}$ Mpc$^{-1}$ and $\Omega_{r0}=0.029$ for $k=.002h$Mpc$^{-1}$ (red) and $k=.01h$Mpc$^{-1}$ (purple). Right side: Evolution of the growth factor for the self-accelerating branch using the values obtained for the background parameters shown in Table \ref{tab:self_acc}, $\Omega_{m0}=0.286$, $H_0=69.08$ Km s$^{-1}$ Mpc$^{-1}$ for $k=0.0005h$Mpc$^{-1}$ (red), $k=0.002h$Mpc$^{-1}$ (green) and $k=0.01h$Mpc$^{-1}$ (red). We also show the evolution of the growth factor  in the $\Lambda$CDM model with the parameter values inferred from Planck $\Omega_{m0}=0.315$ and $H_0=67.4$ Km s$^{-1}$ Mpc$^{-1}$.}
    \label{fig:growthfactor}
\end{figure}

The Friedmann equation for the self-accelerating is obtained from (\ref{ec:friedmann}) replacing $\epsilon=1$ and $\sigma=0$, then
\begin{equation}
\label{Hself}
    H=H_0\left(\sqrt{\frac{\Omega_{m0}}{a^3}+\Omega_{r0}}+\sqrt{\Omega_{r0}}\right),
\end{equation}
As we don't include curvature, the present value of $\Omega_r$, denoted as $\Omega_{r0}$, has to be: 
\begin{equation}
    \Omega_{r0}=\left(\frac{1-\Omega_{m0}}{2}\right)^2,
\end{equation}
hence, in this case, the model parameters are $H_0$, $\Omega_{m0}$ and $M$ for supernovae observations and $H_0$, $\Omega_{m0}$ and $m$ for data of gravitational waves. The priors used are shown in Table \ref{tab:self_acc} and the best-fit values found for the sum of data are: $\Omega_{m0}=0.286\pm0.016$, $H_0=69.08\pm 0.49$ Km Mpc$^{-1}$s$^{-1}$, which differs a little with the estimated value found previously for this model $\Omega_{m0}=0.26^{+0.05}_{-0.04}$  in \cite{lazkoz2007cosmological} and from the values $\Omega_{m0}=0.249\pm0.02$, $\Omega_{r0}=0.1410\pm0.0075$ found in \cite{lombriser2009cosmological}. However the value of $\Omega_{m0}$ is lower than the inferred value from Planck $\Omega_{m0}=0.315\pm0.007$ \cite{aghanim2020planck} assuming a $\Lambda$CDM cosmology. While the value of $H_0$ is similar to the obtained value from Planck, $H_0=67.4\pm0.5$.
\\ The evolution of $\triangle_m$ is obtained by solving (\ref{eqtrianglea}) with $p=4$, $\epsilon=1$ and using (\ref{omegasol}), (\ref{ec:dGdx2}). The evolution of the growth factor $\triangle_m/a$ is shown in Figure \ref{fig:growthfactor} (right side) for different values of $k$, as you can see, the evolution of the growth factor is the same as found in \cite{DGPnumericsolutionscardoso} using the numerical solution.

Then in order to constrain $\sigma_8$ we solve numerically (\ref{eqtrianglea}) using (\ref{ec:dGdx2})
and (\ref{omegasol}) with $\epsilon=1$ for $k=(h/8)Mpc^{-1}$ with initial condition $\triangle_m/a_i$=1, with $a_i=0.01$. For this model, we obtain the value of $\sigma_8=0.913\pm0.032$.


\begin{table}
    \centering

    \begin{tabular}{|c|c|c|c|c|}
        \hline
		Parameter & Prior& SN &GW&SN+GW\\
  \hline
  $\Omega_{m0}$ &(0, 1)& $0.294\pm0.018$&$0.336\pm 0.068$&$0.286\pm0.016$\\
  \hline
  $H_0$ [Km s$^{-1}$ Mpc$^{-1}$]&(66, 74) &$70.26\pm 0.98$&$68.58\pm 0.83$ &$69.08\pm 0.49$\\
  \hline
  M&$(-21,-18)$&$-19.263\pm0.030$&$-$&$-19.302\pm0.014$\\
  \hline
  m&$(0.1,10)$ &$-$&$4.8\pm2.9$ &$5.7\pm 1.9$ \\
  \hline
    \end{tabular}
    \caption{Constraints of the self-accelerating branch.}
    \label{tab:self_acc}
\end{table}

\begin{figure}
    \centering
    \includegraphics[width=8.5cm]{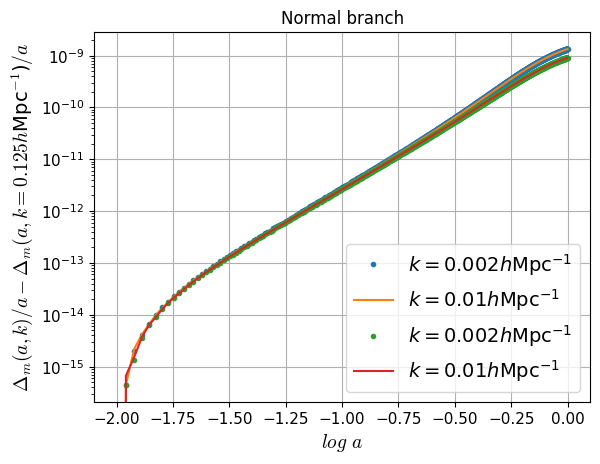}\includegraphics[width=8.5cm]{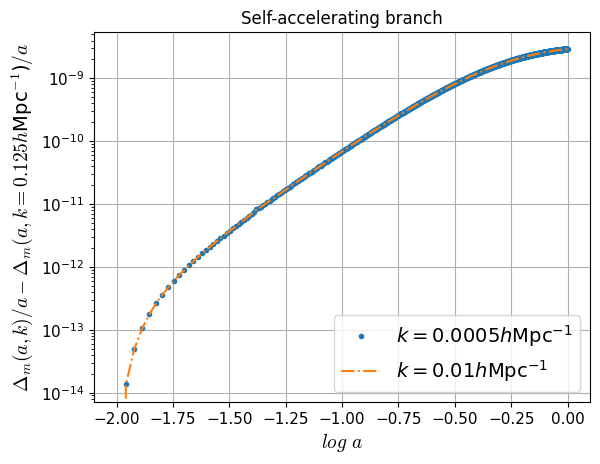}
    \caption{Left side: a) Difference of $\triangle_m(a,k)/a-\triangle_m(a,k=0.125h$Mpc$^{-1}$)$/a$ for different values of $k$ for the normal branch. The difference for these values of $k$ is approximately 0. We use for the background the values shown in Table \ref{tab:nor1}: $\Omega_{m0}=0.546$, $H_0=69.58$ Km s$^{-1}$ Mpc$^{-1}$ and $\Omega_{r0}=0.155$ for $k=0.002h$Mpc$^{-1}$ (blue dot) and $k=0.01h$Mpc$^{-1}$ (orange line). b) And  for $k=0.002h$Mpc$^{-1}$ (green dot) and $k=0.01h$Mpc$^{-1}$ (red line), we use the background values shown in Table \ref{tab:nor2}: $\Omega_{m0}=0.470$, $H_0=69.94$ Km s$^{-1}$ Mpc$^{-1}$ and $\Omega_{r0}=0.029$. Right side: Difference of $\triangle_m(a,k)/a-\triangle_m(a,k=0.125h$Mpc$^{-1}$)$/a$ for different values of $k$ for the self-accelerating branch, using the background parameters shown in Table \ref{tab:self_acc}, $\Omega_{m0}=0.286$, $H_0=69.08$ Km s$^{-1}$ Mpc$^{-1}$ for $k=0.0005h$Mpc$^{-1}$ (blue dot), $k=0.01h$Mpc$^{-1}$ (orange dash-dot line).}
    \label{fig:difference}
\end{figure}


\begin{figure}
    \centering
    \includegraphics[width=12cm]{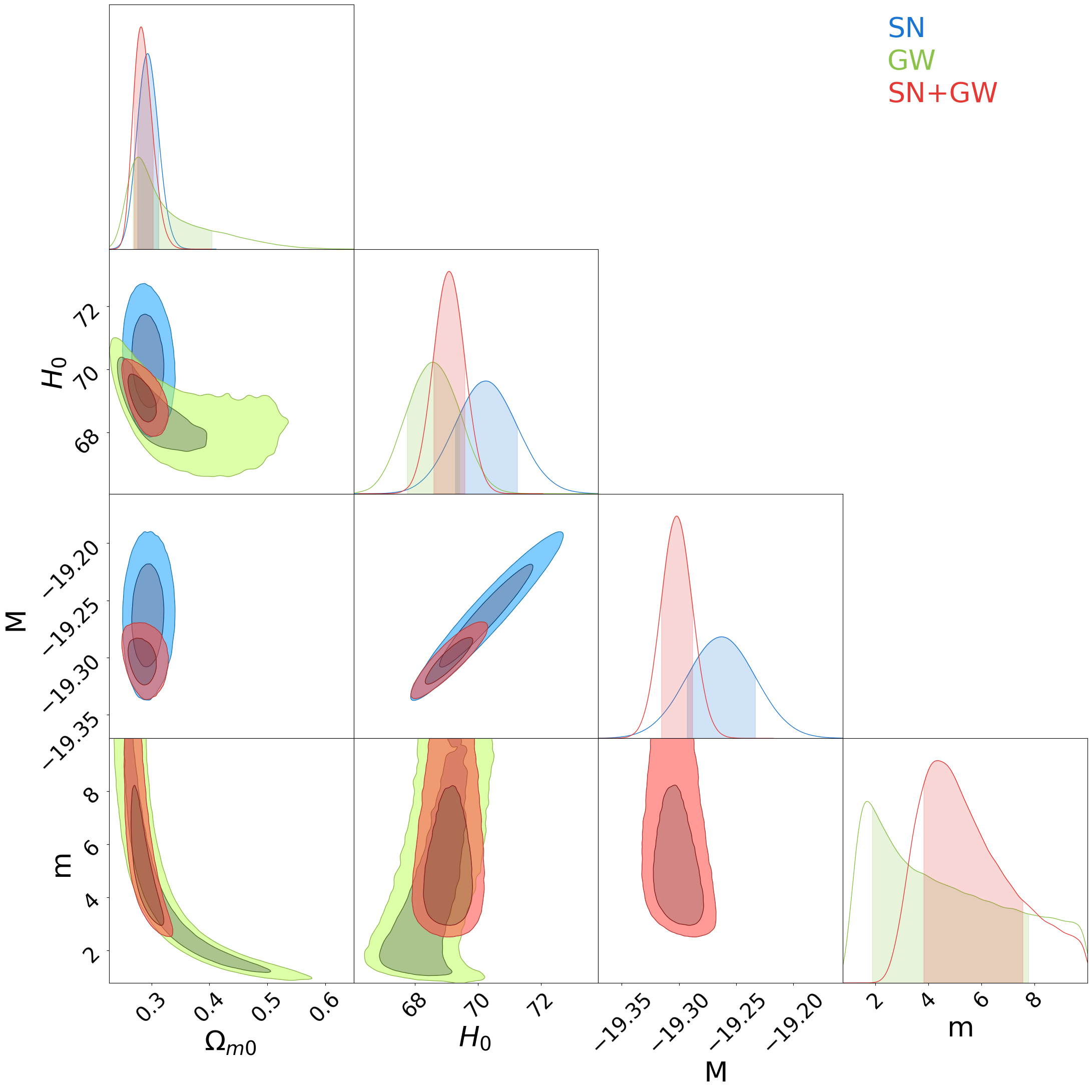}
    \caption{$2\sigma$ C.L. constraints for the background parameters using standard sirens mock data GW (green) and SN Pantheon+ (blue), and the total sample GW+Pantheon+ (red) for the self-accelerating branch using the priors shown in Table \ref{tab:self_acc}.}
    \label{fig:posterior_acelera}
\end{figure}


\begin{figure}
    \centering
     \includegraphics[width=5cm]{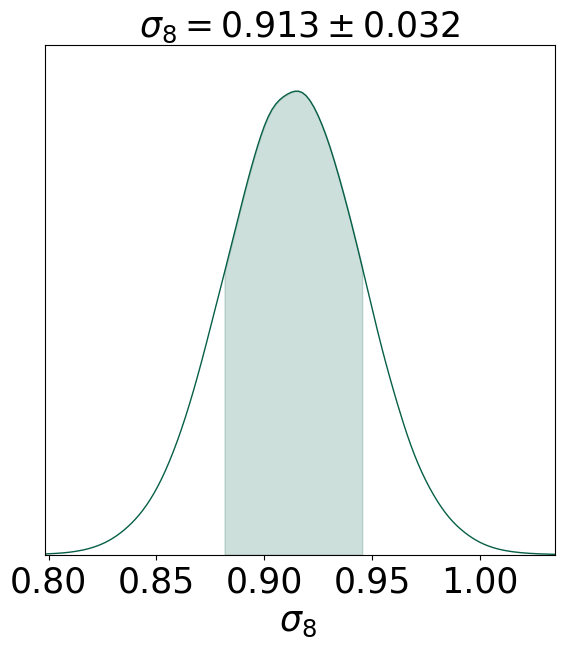}
    \caption{Posterior of $\sigma_8$ for the self-accelerating branch using for the background the best-fit values shown in Table \ref{tab:self_acc}: $\Omega_{m0}=0.286$,  $H_0=69.08$ Km s$^{-1}$ Mpc$^{-1}$. We use a uniform prior for $\sigma_8\epsilon(0,1.2)$.}
    \label{sigma8normal}
\end{figure}


\begin{table}
 \begin{tabular}{|c|c|c|c|c|c|}
        \hline
		Parameter &Normal branch&Self-accelerating branch&\multicolumn{3}{c}{$\Lambda$CDM}\\
        &&&Planck&SH0ES&Kids-1000\\
        \hline
  $\Omega_{m0}$ &$0.470\pm0.028$& $0.286\pm0.016$&$0.315\pm0.007$&$0.334\pm0.018$&$-$\\
  \hline
  $H_0$ [Km s$^{-1}$ Mpc$^{-1}$]& $69.94\pm 0.57$ &$69.08\pm 0.49$&$67.4\pm 0.5$&$73.3\pm1.04$&$-$\\
  \hline
$\sigma_8$&$0.774\pm0.027$&$0.913\pm0.032$&$0.811\pm0.006$&$-$&$0.76^{+0.021}_{-0.023}$\\
  \hline
    \end{tabular}
    \caption{Comparison of the values of the main cosmological parameters.}
    \label{tab:comparison}
\end{table}


\section{Conclusions}
\label{sec:conclusions}

To constrain the value of $\sigma_8$, we study the evolution of density perturbations assuming the scaling solution for the master equation and the boundary condition found in \cite{DGPnumericsolutionscardoso}, then we obtain a second order differential equation for density perturbations that we solve numerically. But before solving the equation for density perturbations, we first constrain the model parameters for the background. This represents an important update since the Pantheon+ catalog enable us to constrain $H_0$ and $M$ independently, for both branches and to the best of our knowledge, such an analysis has not previously been performed for any of these models.\\
Earlier works on self-accelerating branch, such as those by \cite{fang2008challenges} and others, showed that while the DGP model could qualitatively reproduce late-time cosmic acceleration, it faced tensions with combined data sets
including CMB, large scale structure, and supernovae, particularly in fitting the
expansion history and growth of structure simultaneously. These studies typically constrained parameters like the matter density $\Omega_{m0}$ using earlier supernova
compilations combined with BAO and CMB shift parameters, with reported value of $\Omega_{m0}=0.26\pm0.02$, $\Omega_{r0}=0.136\pm0.009$ and Hubble constant $H_0=66\pm2$ Km s$^{-1}$ Mpc$^{-1}$ \cite{fang2008challenges}. The self-accelerating branch, in particular, showed some discrepancies
with weak lensing and large scale structure data.\\
Our analysis improves upon
these earlier results by using the updated Pantheon+ data set which contains many more supernovae
and better-calibrated distances measurements, alongside projected gravitational wave standard
siren data that provide independent measures of the expansion history. These
combined data allow for a more precise Bayesian parameter estimation, not only for background parameters $\Omega_{m0}, H_0$, and $\Omega_{r0}$  but also for the amplitude of matter fluctuations $\sigma_8$ through the evolution
of scalar perturbations.
In summary we find the following:\\
\textbf{Background constraints}
\begin{itemize}
\item \textbf{Normal branch}: we consider two different priors for $\Omega_{r0}$. The first prior yields $\Omega_{m0}=0.546\pm0.039$, $\Omega_{r0}=0.155\pm0.067$, $H_0=69.58\pm0.53$ Km s$^-1$Mpc$^{-1}$ while for a tighter prior   we obtain $\Omega_{m0}=0.470\pm0.028$, $H_0=(69.94\pm0.57)$Km s$^{-1}$ Mpc$^{-1}$, $\Omega_{r0}=0.029$. We conclude that the value of $\Omega_{r0}$ has to be small to obtain a lower value of $\Omega_{m0}$, however the resulting $\Omega_{m0}$ is still higher than that in the $\Lambda$CDM model. For comparison in a flat $\Lambda$CDM cosmology using Pantheon+ and SH0ES data, $\Omega_{m0}=0.334\pm0.018$ and $H_0=(73.6\pm1.1)$ Km s$^{-1}$ Mpc$^{-1}$ \cite{brout2022pantheon+}, while Planck data indicate that $\Omega_{m0}=0.315\pm0.0007$, $H_0=(67.4\pm0.5)$ Km s$^{-1}$ Mpc$^{-1}$  assuming a $\Lambda$CDM cosmology \cite{aghanim2020planck}. At the same time this value of $\Omega_{m0}$ is slightly higher than
some older estimates for DGP model but consistent within uncertainties with prior ranges \cite{rydbeck2007testing}.
\item \textbf{Self-accelerating branch:} $\Omega_{m0}=0.286\pm0.016$ and $H_0=69.08\pm 0.49$ km s$^{-1}$Mpc${^{-1}}$, comparable to but slightly more precise than previous constraints, e.g., $\Omega_{m0}\approx 0.26$ from ESSENCE data \cite{fang2008challenges}, \cite{rydbeck2007testing}.
\end{itemize}
\textbf{Growth of perturbations and $\sigma_8$}.\\
Using the best-fit values for $\Omega_{m0}$ and $H_0$ from Table \ref{tab:nor1} and Table\ref{tab:nor2}, we solve the perturbation equation for $k=(h/8)$Mpc$^{-1}$ and perform the statistical analysis described in Section \ref{sec:fsigma8}. We obtain:
\begin{itemize}
\item \textbf{Normal branch:}\\
$\sigma_8=0.714\pm0.025$ when $\Omega_{m0}=0.546\pm0.039$ and $H_0=(69.58\pm0.53)$ km s$^{-1}$ Mpc$^{-1}$\\ $\sigma_8=0.774\pm0.027$ when $\Omega_{m0}=0.470\pm0.028$ and $H_0=69.94\pm0.57$ km s$^{-1}$ Mpc$^{-1}$.
\item \textbf{Self-accelerating branch:} $\sigma_8=0.913\pm0.032$.
 \end{itemize}
 The growth rate $\Delta_m/a$ with respect to $\log a$ shown in Figure \ref{fig:growthfactor} , confirms the evolution of density perturbations found in \cite{DGPnumericsolutionscardoso}. The deviation from the $\Lambda$CDM model, for the normal branch increases with larger $\Omega_{m0}$. Moreover, for both branches and for different values of $k$, the evolution of $\triangle_m/a$ remains nearly identical, as shown in Figure \ref{fig:growthfactor} and verified in Figure \ref{fig:difference} where the difference $\triangle_m(a,k)/a-\triangle_m(a,k=(h/8)$Mpc$^{-1}$) is approximately zero.\\
 Therefore, we find that $\Omega_{m0}$ for the normal branch is greater than in $\Lambda$CDM, while for the self-accelerating branch it is smaller. In both branches, $H_0$ lies between the CMB (Planck) and SH0ES determinations for $\Lambda$CDM.\\
 The $\sigma_8$ values obtained for both branches differ from that inferred from Planck CMB data for $\Lambda$CDM ($\sigma_8 = 0.811 \pm 0.006$\cite{aghanim2020planck}), and also from large-scale structure measurements ($\sigma_8 = 0.76^{+0.021}_{-0.023}$)~\cite{heymans2021kids}.  
 This comparison suggests that the normal branch better fits the clustering amplitude seen in the matter distribution specially compared to the $\Lambda$CDM model  prediction from Planck CMB data $(\sigma_8\approx 0.811)$ which is known to exhibit some tension with low-redshift structure observations. Therefore, the normal branch may offer a modified gravity explanation that alleviates
the $\sigma_8$ tension present in $\Lambda$CDM. In contrast, the
self-accelerating branch predicts a significantly higher $\sigma_8 = 0.913 \pm 0.032$, which
is inconsistent with the large-scale structure measurements and is thus disfavored
observationally. Beyond the $\sigma_8$ tension, the self-accelerating branch faces additional theoretical challenges such as the presence of ghost instabilities and strong
coupling issues at relatively small scales $(\approx 1000)$ Km, which further question its
physical viability \cite{luty2003strong}. These points imply that while the self-accelerating branch is effectively ruled out as a viable cosmological model due to
both theoretical and observational problems, the normal branch remains as an interesting candidate that could be compatible with current cosmological data and
might help address existing tensions in cosmology. The normal branch, however,
requires additional components (such as brane tension acting like a cosmological
constant) to explain late-time acceleration, thereby resembling $\Lambda$CDM but with
distinctive predictions for structure growth and gravitational wave propagation
that are testable with future data, as it is showed in this work.\\
Table \ref{tab:comparison} summarizes the comparison between the self-accelerating and normal branches with the $\Lambda$CDM model. From these results, we conclude that:
\begin{itemize}
    \item The normal branch of the DGP model remains a viable extension or modification of $\Lambda$CDM, with background parameters and growth factors that are
broadly consistent with current observations, potentially addressing some
tensions present in $\Lambda$CDM.
\item The self-accelerating branch is largely disfavored both observationally and theoretically.
\item The $\Lambda$CDM model continues to provide the best overall fit to current data
but is not free from unresolved tensions, motivating the continued exploration
of alternatives such as the normal DGP branch.
\end{itemize}
Finally, the
discrepancy in $\sigma_8$ estimates supports a cautious rejection of the self-accelerating
branch of the DGP model, while keeping the normal branch as a viable and
promising modified gravity framework. This result highlights the importance of precise
measurements of structure growth and demonstrates the usefulness of the DGP model’s distinct
predictions for testing gravity on cosmological scales. Thus, the DGP model, in its
normal branch formulation, should not be excluded outright and remains worthy of further investigation.\\
In summary, the similarity in parameter values between the normal DGP branch and $\Lambda$CDM underscores the former as a serious candidate for extended cosmological models that modify gravity at large scales.


\section*{Acknowledgements}
MH-M acknowledges financial
support from SECIHTI postdoctoral fellowships. 
CE-R acknowledges the Royal Astronomical Society as FRAS 10147. 


\appendix
\section{Appendix}

In \cite{onbraneworldcedric} it was shown that the scalar perturbations are related to $\Omega$ as:
\begin{eqnarray}
\label{perturbations5d}
    {A}=-\frac{1}{6b}\left[2\Omega''-\frac{n'}{n}\Omega'+\frac{\Lambda_5}{6}\Omega+\frac{1}{n^2}\left(\ddot{\Omega}-\frac{\dot{n}}{n}\dot{\Omega}\right)\right],\nonumber\\
    {A_y}=\frac{1}{nb}\left(\dot{\Omega}'-\frac{n'}{n}\dot{\Omega}\right),\nonumber\\
    \mathcal{R}=\frac{1}{6b}\left(\Omega''-\frac{1}{n^2}\ddot{\Omega}+\frac{\Lambda_5}{3}\Omega+\frac{\dot{n}}{n^3}\dot{\Omega}+\frac{n'}{n}\Omega'\right).
\end{eqnarray}
From the master equation (\ref{mastereq}) we can obtain $\Omega''$ 
\begin{equation}
    \Omega''=\frac{\ddot{\Omega}}{n^2}-\frac{1}{n^2}\left(\frac{\dot{n}}{n}+3\frac{\dot{b}}{b}\right)\dot{\Omega}+\left(3\frac{b'}{b}-\frac{n'}{n}\right)\Omega'+\left(\frac{\Lambda_5}{6}+\frac{k^2}{b^2}\right)\Omega
\end{equation}
and if we replace $\Omega''$ in the set of equations (\ref{perturbations5d}) and evaluating $n$ and $b$ on the brane, we obtain:
\begin{eqnarray}
\label{AR}
    {A}&=&\frac{1}{6a}\left[-3\ddot{\Omega}_b+6H\dot{\Omega}_b+3\epsilon\Omega'_b\left(\frac{\dot{H}}{H}-H\right)-\left(\frac{2k^2}{a^2}+\frac{\Lambda_5}{2}\right)\Omega_b\right],\nonumber\\
\mathcal{R}&=&\frac{1}{6a}\left[3\epsilon H\Omega'_b-3H\dot{\Omega}_b+\left(\frac{\Lambda_5}{2}+\frac{k^2}{a^2}\right)\Omega_b\right].
\end{eqnarray}
where the subscript $b$ indicates that it is being evaluated on the brane at $y=0$, and in this work we consider $\Lambda_5=0$.

In the $5D$ longitudinal gauge, the location of the brane is perturbed and given by
\begin{equation}
\label{bending}
    y=\xi=-r_c(\Phi+\Psi), 
\end{equation}
and the induced metric perturbations on the brane are 
\begin{eqnarray}
\label{inducedpert}
    \Psi&=&A-\epsilon\left(\frac{\dot{H}}{H}+H\right)\xi,\hspace{.5cm}  \Phi=\mathcal{R}-\epsilon H\xi,
\end{eqnarray}
from (\ref{bending}) and (\ref{inducedpert}) it can be found that: 
\begin{eqnarray}
\label{Phi}
\Phi&=&\frac{1}{1-r_c\epsilon(\dot{H}/H+2H)}\left[R\left\lbrace1-\epsilon r_c\left(\frac{\dot{H}}{H}+H\right)\right\rbrace+\epsilon H r_c A\right],\\
\label{Psi}
    \Psi&=&\frac{1}{1-r_c\epsilon(\dot{H}/H+2H)}\left[(1-\epsilon Hr_c)A+r_c\epsilon \mathcal{R}\left(\frac{\dot{H}}{H}+H\right)\right],
\end{eqnarray}
then using (\ref{AR}) and the boundary condition (\ref{boundaryomega}) we can rewrite $\Phi$ and $\Psi$ in terms of $\Omega_b$ given by (\ref{phimaster}).


\section{Scaling solution}
\label{app:scalingsolution}
We assume a scaling solution \cite{largescalesong} for $\Omega$ given by 
\begin{equation}
    \Omega(a,x)=\mathcal{A}a^pG(x)
\end{equation}
such that $\Omega|_{y=0}=\mathcal{A}a^p$, with $x\equiv yH$, and $G|{y=0}=1$.

The causal horizon of the propagation of perturbations through the bulk is given by 
\begin{equation}
    \xi=aH^2\int_{0}^a\frac{da'}{a'^2H(a')^2}
\end{equation}
then $G(x=\xi)=0.$

With this, we can obtain: 
\begin{eqnarray}
\label{domega}
    \dot{\Omega}&=&\mathcal{A}a^pH(pG+xh\frac{dG}{dx})\nonumber\\
    \ddot{\Omega}&=&H^2\mathcal{A}a^p\left[(h^2+hp)G+(2pxh+2xh^2+x\tilde{h})\frac{dG}{dx}+x^2h^2\frac{d^2G}{dx^2}\right],
\end{eqnarray}
where $h=(dH/d\ln a)/H=\frac{a}{H}\frac{dH}{da}$ and $\tilde{h}=\frac{dh}{d\ln a}$.
And evaluating at $y=0$, it can be found that: 
 \begin{eqnarray}
 \label{ec:dotomega}
\Omega|_{y=0}&=&\mathcal{A}a^p\nonumber\\
    \dot{\Omega}|_{y=0}&=&\mathcal{A}pa^pH\nonumber\\
    \ddot{\Omega}|_{y=0}&=&H^2\mathcal{A}a^p(h^2+hp)=\mathcal{A}H^2a^p\left[\frac{a^2}{H^2}\left(\frac{dH}{da}\right)^2+\frac{p}{H}\frac{dH}{da}\right].
\end{eqnarray}
Then replacing (\ref{domega}) in (\ref{mastereq}) and using (\ref{nyb}) we can find a differential equation for $G(x)$ for the normal branch given by: 
\begin{equation}
    A(x)\frac{d^2G}{dx^2}+B(x)\frac{dG}{dx}+C(x)G=0,
\end{equation}
where 
\begin{eqnarray}
A(x)&=&(1-x)(1-x-2hx),\nonumber\\
B(x)&=&-2x(hp+1)+2-h+\frac{(x^2-x)(h^2+h'+h)}{1-x(h+1)},\nonumber\\
C(x)&=&-p^2-hp-\frac{xp(h'+h^2+h)}{1-x(h+1)}+\frac{3p(1-x-xh)}{1-x}-\frac{[1-x(1+h)]^2}{(1-x)^2}\frac{k^2}{a^2H^2},
\end{eqnarray}
that is a simplified equation of the version found in \cite{largescalesong}.
While the accelerated branch is:
\begin{eqnarray}
    A(x)&=&(1+x)(1+x(1+2h)),\nonumber\\
    B(x)&=&-2x(hp+1)-2+h-\frac{(x^2+x)(h^2+h'+h)}{1+x(h+1)},\nonumber\\
    C(x)&=&-p^2-hp+\frac{xp(h'+h^2+h)}{1+x(h+1)}+\frac{3p(1+x+xh)}{1+x}-\frac{[1+x(h+1)]^2}{(1+x)^2}\frac{k^2}{a^2H^2},
\end{eqnarray}
and this is equivalent to the equation found in \cite{nearhorizon}.\\
And according to \cite{DGPnumericsolutionscardoso} it can be found a boundary condition for $\Omega$ at $y=0$ which is: 
\begin{eqnarray}
\label{boundary}
    (\partial_y\Omega)_b=-\frac{\epsilon\gamma_1}{2H}\ddot{\Omega}_b+\frac{9\epsilon\gamma_3}{4}\dot{\Omega}_b-\left(\frac{3\epsilon\gamma_3k^2}{4Ha^2}+\frac{H\gamma_4}{4}\right)\Omega_b+\frac{3\epsilon r_c\kappa_4^2\rho a^3\gamma_4}{2k^2}\triangle,
\end{eqnarray}
where $\gamma_1,\gamma_2,\gamma_3,\gamma_4$ are given by:
\begin{eqnarray}
\label{ec:gammas}
\gamma_1&=&\frac{2\epsilon Hr_c}{2\epsilon H r_c-1}=\frac{\epsilon H}{\epsilon H-H_0\sqrt{\Omega_{r0}}},\nonumber\\
    \gamma_2&=&\frac{2\epsilon r_c(\dot{H}-H^2+2\epsilon H^3 r_c)}{H(2\epsilon Hr_c-1)^2}=\frac{\epsilon H_0\sqrt{\Omega_{r0}}\left(a\frac{dH}{da}-H\right)+H^2}{(\epsilon H-H_0\sqrt{\Omega_{r0}})^2},\nonumber\\
\gamma_3&=&\frac{4\epsilon r_c(2\epsilon r_c\dot{H}-3H+6\epsilon H^2 r_c)}{9(2\epsilon H r_c-1)^2}=\frac{2aH\frac{dH}{da}+6H^2-6H\epsilon H_0\sqrt{\Omega_{r0}}}{9(\epsilon H-H_0\sqrt{\Omega_{r0}})^2},\nonumber\\
\gamma_4&=&\frac{4\epsilon(\epsilon r_c\dot{H}-H+2\epsilon H^2 r_c)}{3H(2\epsilon Hr_c-1)^2}=\frac{H_0\sqrt{\Omega_{r0}}}{(\epsilon H-H_0\sqrt{\Omega_{r0})^2}}\left(\frac{2}{3}a\frac{dH}{da}-\frac{4}{3}H_0\sqrt{\Omega_{r0}}\epsilon+\frac{4}{3}H\right).
\end{eqnarray}
Then if we consider $\Omega=\mathcal{A}a^pG$ and using (\ref{ec:dotomega}), then the boundary condition (\ref{boundaryomega}) can be rewritten as: 
\begin{equation}
\label{ec:dGdx}
    \frac{dG}{dx}|_{y=0}=-\frac{\epsilon\gamma1}{2}(h^2+hp)+\frac{9\epsilon\gamma_3}{4}p-\frac{3\epsilon\gamma_3k^2}{4H^2a^2}-\frac{H\gamma_4}{4}+\frac{3\epsilon r_c\kappa_4^2\rho a^3\gamma_4}{2k^2}\frac{\triangle}{\mathcal{A}a^p H},
\end{equation}
where we have used $\Omega_b=\mathcal{A}a^p G(x=0)=\mathcal{A}a^p$.\\

\bibliographystyle{unsrt}
\bibliography{bibliography}
\end{document}